\documentclass[nonacm, sigconf]{acmart}
\usepackage{shellesc}
\usepackage{algorithmic}
\usepackage{graphicx}
\usepackage{textcomp}
\usepackage{xcolor}
\usepackage{svg}
\def\BibTeX{{\rm B\kern-.05em{\sc i\kern-.025em b}\kern-.08em
    T\kern-.1667em\lower.7ex\hbox{E}\kern-.125emX}}
    
\usepackage{float}    
\usepackage{bera}
\usepackage{listings,mdframed}
\usepackage{xcolor}
\usepackage{tikz}
\usepackage{pgfplots}
\usepackage{subcaption}
\usepackage{float}
\pgfplotsset{compat=1.12}
\usepackage{pythonhighlight}
\usepackage[frozencache=true,cachedir=.]{minted}
\definecolor{comment-text-color}{rgb}{0.15, 0.8, 0.45}
\lstdefinelanguage{json}{
    basicstyle=\normalfont\ttfamily\footnotesize,
    numbers=left,
    numberstyle=\scriptsize,
    stepnumber=1,
    numbersep=8pt,
    showstringspaces=false,
    breaklines=true,
    frame=lines,
    backgroundcolor=\color{background},
    literate=
     *{0}{{{\color{numb}0}}}{1}
      {1}{{{\color{numb}1}}}{1}
      {2}{{{\color{numb}2}}}{1}
      {3}{{{\color{numb}3}}}{1}
      {4}{{{\color{numb}4}}}{1}
      {5}{{{\color{numb}5}}}{1}
      {6}{{{\color{numb}6}}}{1}
      {7}{{{\color{numb}7}}}{1}
      {8}{{{\color{numb}8}}}{1}
      {9}{{{\color{numb}9}}}{1}
      {:}{{{\color{punct}{:}}}}{1}
      {,}{{{\color{punct}{,}}}}{1}
      {\{}{{{\color{delim}{\{}}}}{1}
      {\}}{{{\color{delim}{\}}}}}{1}
      {[}{{{\color{delim}{[}}}}{1}
      {]}{{{\color{delim}{]}}}}{1},
}

\usepackage{xcolor}
\usepackage{listings}
\usepackage{enumerate}
\usepackage{siunitx}
\usepackage{booktabs}
\usepackage{hyperref}
\usepackage[htt]{hyphenat}
\hypersetup{
     colorlinks   = true,
     citecolor    = blue
}

\definecolor{red}{rgb}{1,0.,0}

\usepackage[compatibility=false]{caption}
\usepackage{tcolorbox}
\setcopyright{none}
\renewcommand\footnotetextcopyrightpermission[1]{} 
\usepackage{booktabs} 

\begin{document}

\newcommand{\tsh}[1]{{#1}}
\newcommand{\tshc}[1]{{\color{blue}[#1]}}
\newcommand{\tsc}[1]{\textsuperscript{#1}} 

\title{QuaSiMo: A Composable Library to Program Hybrid Workflows for Quantum Simulation}
\author{Thien Nguyen$^{1,5}$, Lindsay Bassman$^{2}$, Phillip C.~Lotshaw$^{4,5}$, Dmitry Lyakh$^{3,5}$}
\author{Alexander McCaskey$^{1,5}$, Vicente Leyton-Ortega$^{4,5}$, Raphael Pooser$^{4,5}$}
\author{Wael Elwasif$^{1,5}$, Travis S.~Humble$^{4,5}$, Wibe A. de Jong$^{2}$}
\affiliation{
  \institution{1. Computer Science \& Mathematics Division, Oak Ridge National Laboratory, Oak Ridge, TN, 37831, USA}
  \institution{2. Computational Research Division, Lawrence Berkeley National Laboratory, Berkeley, CA, 94720, USA}
  \institution{3. National Center for Computational Sciences, Oak Ridge National Laboratory, Oak Ridge, TN, 37831, USA}
  \institution{4. Computer Science and Engineering Division, Oak Ridge National Laboratory, Oak Ridge, TN, 37831, USA}
  \institution{5. Quantum Computing Institute, Oak Ridge National Laboratory, Oak Ridge, TN, 37831, USA}
}

\begin{abstract}
We present a composable design scheme for the development of hybrid quantum/classical algorithms and workflows for applications of quantum simulation. Our object-oriented approach is based on constructing an expressive set of common data structures and methods that enable programming of a broad variety of complex hybrid quantum simulation applications. The abstract core of our scheme is distilled from the analysis of the current quantum simulation algorithms. Subsequently, it allows a synthesis of new hybrid algorithms and workflows via the extension, specialization, and dynamic customization of the abstract core classes defined by our design. We implement our design scheme using the hardware-agnostic programming language QCOR into the \texttt{QuaSiMo} library. To validate our implementation, we test and show its utility on commercial quantum processors from IBM and Rigetti, running some prototypical quantum simulations.
\end{abstract}

\thanks{This manuscript has been authored by UT-Battelle, LLC under Contract No. DE-AC05-00OR22725 and Lawrence Berkeley National Laboratory under Contract No. DE-AC02-05CH11231 with the U.S. Department of Energy. The United States Government retains and the publisher, by accepting the article for publication, acknowledges that the United States Government retains a non-exclusive, paid-up, irrevocable, world-wide license to publish or reproduce the published form of this manuscript, or allow others to do so, for United States Government purposes. The Department of Energy will provide public access to these results of federally sponsored research in accordance with the DOE Public Access Plan. (http://energy.gov/downloads/doe-public-access-plan).}

\maketitle
\pagestyle{plain}
\section{Introduction}

Quantum simulation is an important use case of quantum computing for scientific computing applications. Whereas numerical calculations of quantum dynamics and structure are staples of modern scientific computing, quantum simulation represents the analogous computation based on the principles of quantum physics. Specific applications are wide-ranging and include calculations of electronic structure \cite{aspuru2005simulated,whitfield2011simulation,cao2019quantum, mcardle2020quantum}, scattering \cite{yeter2020scattering}, dissociation \cite{o2016scalable}, thermal rate constants \cite{lidar1999calculating}, materials dynamics \cite{bassman2020towards}, and response functions \cite{kosugi2020linear}.
\par
Presently, this diversity of quantum simulation applications is being explored with quantum computing despite the limitations on the fidelity and capacity of quantum hardware \cite{PhysRevLett.124.230501,cirstoiu2020variational,google2020hartree}. These applications are tailored to such limitations by designing algorithms that can be tuned and optimized in the presence of noise or model representations that can be reduced in dimensionality.  Examples include variational methods such as the variational quantum eigensolver (VQE) \cite{mcclean2016theory,romero2018strategies,grimsley2019adaptive,tang2019qubit}, quantum approximate optimization algorithm (QAOA)~\cite{farhi2014quantum}, quantum imaginary time evolution (QITE) \cite{motta2019determining}, and quantum machine learning (QML) among others. 
\par
The varied use of quantum simulation raises concerns for efficient and effective programming of these applications. The current diversity in quantum computing hardware and low-level, hardware-specific languages imposes a significant burden on the application user. For instance, IBM provides Aqua~\cite{Qiskit-Aqua}, which is part of the Qiskit framework, targeting high-level quantum applications such as chemistry and finance. The emphasis of Aqua is on providing robust implementations of quantum algorithms, yet the concept of reusable and extensible workflows, especially for quantum simulations, is not formally supported. The user usually has to implement custom workflows from lower-level constructs, such as circuits and operators, available in Qiskit Terra~\cite{Qiskit-Terra}. Similarly, Tequila~\cite{kottmann2020tequila} is another Python library that provides commonly-used functionalities for quick prototyping of variational-type quantum algorithms. Orquestra~\cite{Orquestra}, a commercially-available solution from Zapata, on the other hand, orchestrates quantum application workflows as black-box phases, thus requires users to provide implementations for each phase.
\par
The lack of a  common workflow for applications of quantum simulation hinders broader progress in testing and evaluation of such hardware. A common, reusable and extensible programming workflow for quantum simulation would enable broader adoption of these applications and support more robust testing by the quantum computing community.

\par 
In this contribution, we address development of common workflows to unify applications of quantum simulation. Our approach constructs common data structures and methods to program varying quantum simulation applications, and we leverage the hardware-agnostic language QCOR and programming framework XACC to implement these ideas. We demonstrate these methods with example applications from materials science and chemistry, and we discuss how to extend these workflows to experimental validation of quantum computation advantage, in which numerical simulations can benchmark programs for small-sized models \cite{kandala2017hardware,hempel2018quantum,mccaskey2019quantum,google2020hartree,yeter2020practical}. 

\par
Upon release of the \texttt{QuaSiMo} library, we became aware of the additional application modules available to the Qiskit framework which share a number of features with our proposed composable workflow design. In particular, the Qiskit Nature module~\cite{qiskit-nature} introduces new concepts to model and solve quantum simulation problems, and reiterates the need for modular, domain-specific, and workflow-driven quantum application builders, which the \texttt{QuaSiMo} work addresses. The key differentiators for our work lie in the performance and extensibility of the implementation. Since \texttt{QuaSiMo} is developed on top of the C++ QCOR compiler infrastructure, it can take advantage of optimal performance for classical computing components of the workflow, such as circuit construction~\cite{qcor2} or post-processing. Moreover, the plugin-based extension model of \texttt{QuaSiMo} is distinct from that of Qiskit Nature. Rather than requiring new extensions being imported, \texttt{QuaSiMo} puts forward a common interface for all of its extension points~\cite{osgi}, and therefore enables the development of portable user applications w.r.t. the underlying library implementation.  

\begin{figure*}[t!]
    \centering
    \includegraphics[width=\textwidth]{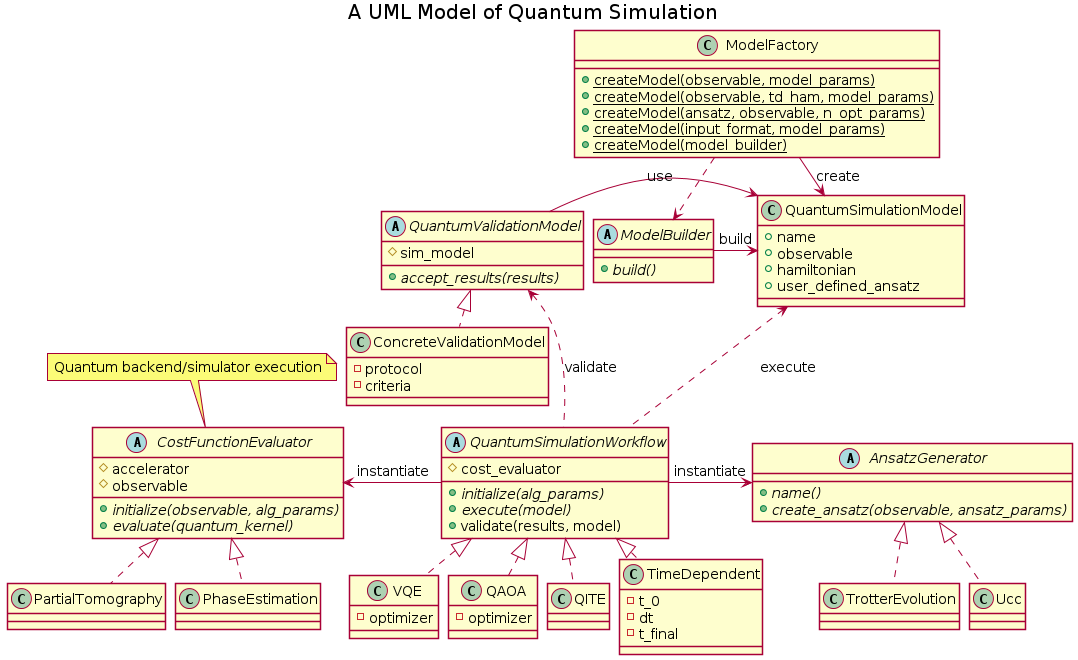}
    \caption{The class UML diagram of the quantum simulation application. The fully typed version is provided separately (see~\cite{paper_repo}).}
    \label{fig:uml}
\end{figure*}

\section{Software Architecture}
Cloud-based access to quantum computing naturally differentiates programming into conventional and quantum tasks \cite{britt2017high,xacc1}. The resulting hybrid execution model yields a loosely integrated computing system by which common methods have emerged for programming and data flow. We emphasize this concept of workflow to organize programming applications for quantum simulation.

Figure~\ref{fig:uml} shows the blueprint of our Quantum Simulation Modeling (\texttt{QuaSiMo}) library. The programming workflow is defined by a \texttt{QuantumSimulationWorkflow} concept which encapsulates the hybrid quantum-classical procedures pertinent to a quantum simulation, e.g., VQE, QAOA, or dynamical quantum simulation. A quantum simulation workflow exposes an \texttt{execute} method taking as input a \texttt{QuantumSimulationModel} object representing the quantum model that needs to be simulated. This model captures quantum mechanical observables, such as energy, spin magnetization, etc., that we want the workflow to solve or simulate for. In addition, information about the system Hamiltonian, if different from the observable operator of interest, and customized initial quantum state preparation can also be specified in the \texttt{QuantumSimulationModel}.

By separating the quantum simulation model from the simulation workflow, our object-oriented design allows the concrete simulation workflow to simulate rather generic quantum models. This design leverages the \texttt{ModelFactory} utility, implementing the object-oriented factory method pattern. A broad variety of input mechanisms, such as those provided by the QCOR infrastructure or based on custom interoperability wrappers for quantum-chemistry software, can thus be covered by a single customizable polymorphic model. For additional flexibility, the last \texttt{createModel} factory method overload accepts a polymorphic builder interface \texttt{ModelBuilder} the implementations of which can build arbitrarily composed \texttt{QuantumSimulationModel} objects.

\texttt{QuantumSimulationWorkflow} is the main extension point of our \texttt{QuaSiMo} library. Built upon the \texttt{CppMicroServices} framework conforming to the Open Services Gateway Initiative (OSGi) standard~\cite{marples2001open}, \texttt{QuaSiMo} allows implementation of a new quantum workflow as a plugin loadable at runtime. At the time of this writing, we have developed the \texttt{QuantumSimulationWorkflow} plugins for the VQE, QAOA, QITE, and time-dependent simulation algorithms, as depicted in Fig.~\ref{fig:uml}. All these plugins are implemented in the QCOR language~\cite{qcor1, qcor2} using the externally-provided library routines.

At its core, a hybrid quantum-classical workflow is a procedural description of the quantum circuit composition, pre-processing, execution (on hardware or simulators), and post-processing. To facilitate modularity and reusability in workflow development, we put forward two concepts, \texttt{AnsatzGenerator} and \texttt{CostFunctionEvaluator}. \texttt{AnsatzGenerator} is a helper utility used to generate quantum circuits based on a predefined method such as the Trotter decomposition~\cite{trotter1959product,suzuki1976generalized} or the unitary coupled-cluster (UCC) ansatz~\cite{barkoutsos2018quantum}. \texttt{CostFunctionEvaluator} automates the process of calculating the expectation value of an observable operator. For example, a common approach is to use the partial state tomography method of adding change-of-basis gates to compute the operator expectation value in the Z basis. Given the \texttt{CostFunctionEvaluator} interface, quantum workflow instances can abstract away the quantum backend execution and the corresponding post-processing of the results. This functional decomposition is particularly advantageous in the NISQ regime since one can easily integrate the noise-mitigation techniques, e.g., the verified quantum phase estimation protocol~\cite{o2020error}, into the \texttt{QuaSiMo} library, which can then be used interchangeably by all existing workflows.

Finally, our abstract \texttt{QuantumSimulationWorkflow} class also exposes a public \texttt{validate} method accepting a variety of concrete implementations of the abstract \texttt{QuantumValidationModel} class via a polymorphic interface. Given the quantum simulation results produced by the \texttt{execute} method of \texttt{QuantumSimulationWorkflow}, the concrete implementations of \texttt{QuantumValidationModel} must implement its \texttt{accept\_results} method based on different validation protocols and acceptance criteria.
For example, the acceptance criteria can consist of distance measures of the results from previously validated values, or from the results of validated simulators. The measure may also be taken relative to experimentally obtained data, which, with sufficient error analysis to bound confidence in its accuracy, can serve as a ground truth for validation. A more concrete example in a NISQ workflow includes the use of the \texttt{QuantumSimulationWorkflow} class to instantiate a variational quantum eigensolver simulator, followed by the use of \texttt{validate} to instantiate a state vector simulator. Results from both simulators can be passed to the \texttt{QuantumValidationModel} \texttt{accept\_results} method which evaluates a distance measure method and optionally calls a decision method which returns a binary answer. Other acceptance criteria include evaluation of formulae with input data, application of curve fits, and user-defined criteria provided in the concrete implementation of the abstract \texttt{QuantumValidationModel} class. The validation workflow relies on the modular architecture of our approach, which effectively means that writing custom validation methods and constructing user-defined validation workflows is achieved by extending the abstract \texttt{QuantumValidationModel} class.

In our opinion, the proposed object-oriented design is well-suited to serve as a pattern for implementing diverse hybrid quantum-classical simulation algorithms and workflows which can then be aggregated inside a library under a unified object-oriented interface. Importantly, our standardized polymorphic design with a clear separation of concerns and multiple extension points provides a high level of composability to developers interested in implementing rather complex quantum simulation workflows.

\par
\section{Testing and Evaluation}

Our implementation of the programming workflow for applications of quantum simulation is available online \cite{qcor_github}. We have tested this implementation against several of the original use cases to validate the correctness of the implementation and to evaluate performance considerations.

\subsection{Dynamical Simulation}
\par
As a first sample use case, we consider a non-equilibrium dynamics simulation of the Heisenberg model in the form of a quantum quench.  A quench of a quantum system is generally carried out by initializing the system in the ground state of some initial Hamiltonian, $H_i$, and then evolving the system through time under a final Hamiltonian, $H_f$.  Here, we demonstrate a simulation of a quantum quench of a one-dimensional (1D) antiferromagnetic (AF) Heisenberg model using the QCOR library to design and execute the quantum circuits.

\begin{figure}[h]
\centering
\begin{tikzpicture}[scale=1.0][domain=0:8]
\begin{axis}[
      cycle list name=exotic,
      legend columns=3,
      xmin = 0, xmax = 100,
      xlabel = {Simulation Timestep}, 
      ylabel = {Staggered Magnetization ($\langle m_s \rangle$)}, 
      y label style={at={(axis description cs:-0.075,0.55)},anchor=south},
      title = {Dynamics of the staggered magnetization}]
\addplot
table[x=x, y=y] {
x   y
0	1
1	0.964846
2	0.863205
3	0.706053
4	0.510195
5	0.296229
6	0.086078
7	-0.0996048
8	-0.243825
9	-0.335134
10	-0.368779
11	-0.347002
12	-0.278445
13	-0.176786
14	-0.0587828
15	0.0579956
16	0.15745
17	0.226945
18	0.258767
19	0.250869
20	0.206832
21	0.135075
22	0.0474436
23	-0.0425946
24	-0.12191
25	-0.179534
26	-0.208102
27	-0.204743
28	-0.171293
29	-0.11384
30	-0.0416791
31	0.0341506
32	0.102534
33	0.153904
34	0.181594
35	0.182742
36	0.158619
37	0.11436
38	0.0581543
39	1.25E-05
40	-0.0496847
41	-0.0816592
42	-0.0890126
43	-0.0681363
44	-0.0191376
45	0.0542617
46	0.145319
47	0.245254
48	0.344466
49	0.433797
50	0.505661
51	0.554881
52	0.57913
53	0.578926
54	0.557219
55	0.518655
56	0.468649
57	0.412436
58	0.354252
59	0.296781
60	0.240956
61	0.186126
62	0.130557
63	0.0721707
64	0.00937177
65	-0.058184
66	-0.129038
67	-0.199711
68	-0.264873
69	-0.317924
70	-0.351892
71	-0.360532
72	-0.33947
73	-0.287185
74	-0.205671
75	-0.100648
76	0.0187628
77	0.140876
78	0.252724
79	0.341624
80	0.396826
81	0.411013
82	0.381448
83	0.31057
84	0.205952
85	0.0795804
86	-0.0534667
87	-0.176795
88	-0.274696
89	-0.334125
90	-0.346382
91	-0.308282
92	-0.222628
93	-0.0979328
94	0.0526027
95	0.212721
96	0.365145
97	0.493755
98	0.585617
99	0.632594
100	0.632322
      };
\addlegendentry{$g = 0.0$}
\addplot
table[x=x, y=y] {
x   y
0	1
1	0.964829
2	0.863211
3	0.706312
4	0.511173
5	0.298569
6	0.090454
7	-0.092698
8	-0.23428
9	-0.323384
10	-0.355845
11	-0.334405
12	-0.26799
13	-0.170243
14	-0.057528
15	0.0533137
16	0.147143
17	0.212359
18	0.242167
19	0.235136
20	0.195003
21	0.129808
22	0.0504964
23	-0.0307822
24	-0.102428
25	-0.154961
26	-0.18219
27	-0.181861
28	-0.155709
29	-0.108948
30	-0.0493136
31	0.0141969
32	0.0726858
33	0.118619
34	0.146826
35	0.155109
36	0.144403
37	0.118484
38	0.0833038
39	0.0460603
40	0.0141456
41	-0.00587055
42	-0.00909358
43	0.00711135
44	0.0428605
45	0.0958285
46	0.161625
47	0.234399
48	0.307578
49	0.374629
50	0.429765
51	0.468489
52	0.487948
53	0.487073
54	0.4665
55	0.428333
56	0.375783
57	0.312761
58	0.243475
59	0.17208
60	0.102416
61	0.037827
62	-0.0189305
63	-0.0657276
64	-0.101067
65	-0.124077
66	-0.134512
67	-0.132761
68	-0.119848
69	-0.0974279
70	-0.0677342
71	-0.0334829
72	0.00229027
73	0.0364494
74	0.0660199
75	0.088465
76	0.101943
77	0.105507
78	0.0992222
79	0.0841747
80	0.0623672
81	0.0365133
82	0.00975125
83	-0.0146876
84	-0.0338148
85	-0.0451848
86	-0.0471252
87	-0.0388726
88	-0.0206045
89	0.00662835
90	0.0410521
91	0.0803794
92	0.122048
93	0.163457
94	0.202186
95	0.236161
96	0.263774
97	0.283948
98	0.296145
99	0.300338
100	0.296953
};
\addlegendentry{$g = 0.25$}
\addplot
table[x=x, y=y] {
x   y
0	1
1	0.966006
2	0.88355
3	0.796277
4	0.740141
5	0.721284
6	0.71976
7	0.711992
8	0.690494
9	0.66548
10	0.65122
11	0.652442
12	0.662581
13	0.672412
14	0.678626
15	0.684549
16	0.694352
17	0.707724
18	0.719735
19	0.724608
20	0.719164
21	0.704155
22	0.684618
23	0.669538
24	0.668658
25	0.685702
26	0.712362
27	0.730084
28	0.721788
29	0.686125
30	0.641723
31	0.615327
32	0.621779
33	0.65239
34	0.682175
35	0.690198
36	0.675539
37	0.655448
38	0.648609
39	0.659763
40	0.67938
41	0.696676
42	0.711543
43	0.73316
44	0.767325
45	0.806173
46	0.831329
47	0.82793
48	0.796444
49	0.752079
50	0.713585
51	0.69186
52	0.68649
53	0.689897
54	0.693396
55	0.691162
56	0.68248
57	0.673568
58	0.676968
59	0.705122
60	0.759544
61	0.824109
62	0.871001
63	0.878083
64	0.844793
65	0.79255
66	0.74818
67	0.724431
68	0.714473
69	0.703923
70	0.687343
71	0.67285
72	0.671117
73	0.68109
74	0.688004
75	0.676634
76	0.646997
77	0.61692
78	0.608261
79	0.62875
80	0.665318
81	0.694372
82	0.69988
83	0.684289
84	0.664697
85	0.65928
86	0.675466
87	0.707323
88	0.741064
89	0.762888
90	0.76492
91	0.748123
92	0.721865
93	0.699362
94	0.689977
95	0.692926
96	0.698088
97	0.694743
98	0.681339
99	0.66699
100	0.66256
      };
\addlegendentry{$g = 4.0$}
\end{axis}
\end{tikzpicture}
\caption{Simulation results of staggered magnetization for an Heisenberg model with nine spins after a quantum quench. The Trotter step size ($dt$) is 0.05.}
\label{fig:stag_mag}
\end{figure}
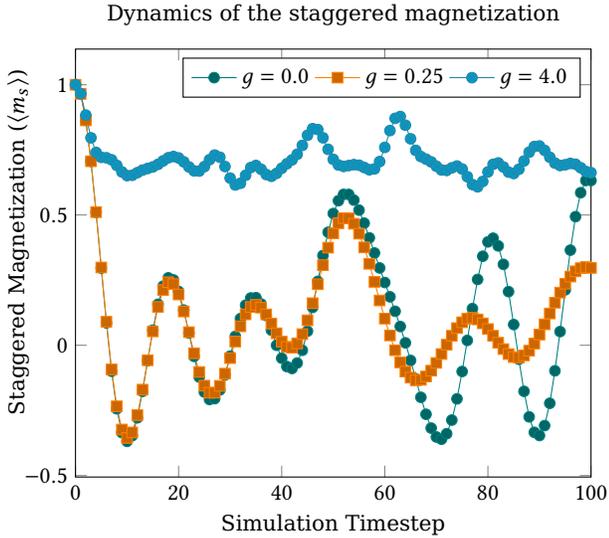

\par
Our AF Heisenberg Hamiltonian of interest is given by 
\begin{equation}
H = J\sum_{\langle i,j \rangle}\{\sigma_{i}^{x}\sigma_{j}^{x} + \sigma_{i}^{y}\sigma_{j}^{y} + g\sigma_{i}^{z}\sigma_{j}^{z} \}
\label{AF_Heisenberg}
\end{equation}
where $J > 0$ gives the strength of the exchange couplings between nearest neighbor spins pairs $\langle i,j \rangle$, $g>0$ defines the anisotropy in the system, and $\sigma_i^\alpha$ is the $\alpha$-th Pauli operator acting on qubit $i$.  We choose our initial Hamiltonian to be the Hamiltonian in equation \ref{AF_Heisenberg} in the limit of $g \rightarrow \infty$. Thus, setting $J=1$, $H_i = C \sum\sigma_{i}^{z}\sigma_{i+1}^{z}$, where $C$ is an arbitrarily large constant.  The ground state of $H_i$ is the N\'eel state, given by $|\psi_0\rangle = |\uparrow \downarrow \uparrow ... \downarrow\rangle$, which is simple to prepare on the quantum computer.  We choose our final Hamiltonian to have a finite, positive value of $g$, so $H_f = \sum_i\{\sigma_{i}^{x}\sigma_{i+1}^{x} + \sigma_{i}^{y}\sigma_{i+1}^{y} + g\sigma_{i}^{z}\sigma_{i+1}^{z}\}$.  Our observable of interest is the staggered magnetization \cite{barmettler2010quantum}, which is related to the AF order parameter and is defined as
\begin{equation}
m_s(t) = \frac{1}{N}\sum_i (-1)^i \langle\sigma_{i}^{z}(t)\rangle
\label{order_param}
\end{equation}
where $N$ is the number of spins in the system.

\begin{figure}[ht!]
\centering
\begin{minted}[frame=single,framesep=5pt]{c++}
using namespace QuaSiMo;
// AF Heisenberg model
auto problemModel = ModelFactory::createModel(
    "Heisenberg", {{"Jx", 1.0},
                  {"Jy", 1.0},
                  // Jz == g parameter
                  {"Jz", g},
                  // No external field
                  {"h_ext", 0.0},
                  {"num_spins", n_spins},
                  {"initial_spins", 
                  initial_spins},
                  {"observable", 
                  "staggered_magnetization"}});
// Time-dependent simulation workflow
auto workflow = 
    std::make_shared<TimeDependentWorkflow>();
workflow->initialize({{"dt", dt}, 
                     {"steps", n_steps}});
// Execute the workflow
auto result = workflow->execute(problemModel);
// Get the observable values 
// (staggered magnetization)
auto obsVals = 
     result.get<std::vector<double>>("exp-vals");
\end{minted}
\caption{Defining the AF Heisenberg problem model and simulating its dynamics with \texttt{QuaSiMo}. In this example, \texttt{g} is the anisotropy parameter, as shown in equation \ref{AF_Heisenberg}, and \texttt{n\_spins} is the number of spins/qubits. \texttt{initial\_spins} is an array of 0 or 1 values denoting the initial spin state. \texttt{initial\_spins} was initialized (not shown here) to a vector of alternating 0 and 1 values (N\'eel state). \texttt{dt} and \texttt{n\_steps} are Trotter step size and number of steps, respectively.}
\label{fig:codesnippet}
\end{figure}

\par
Fig.~\ref{fig:stag_mag} shows sample results for $N = 9$ spins for a three different values for $g$ in $H_f$.  The qualitatively different behaviours of the staggered magnetization after the quench for $g<1$ and $g>1$ are apparent, and agree with previous studies \cite{barmettler2010quantum}. We present a listing of the code expressing this implementation in Fig.~\ref{fig:codesnippet}.

\par
We develop \texttt{QuaSiMo} on top of the QCOR infrastructure, as shown in Fig.~\ref{fig:uml}; thus, any quantum simulation workflows constructed in \texttt{QuaSiMo} are retargetable to a broad range of quantum backends. The results that we have demonstrated in Fig.~\ref{fig:stag_mag} are from a simulator backend. The same code as shown in Fig.~\ref{fig:codesnippet} can also be recompiled with a \texttt{-qpu} flag to target a cloud-based quantum processor, such as those available in the IBMQ network.

Currently available quantum processors, known as noisy intermediate-scale quantum (NISQ) computers \cite{preskill2018quantum}, have relatively high gate-error rates and small qubit decoherence times, which limit the depth of quantum circuits that can be executed with high-fidelity.  As a result, long-time dynamic simulations are challenging for NISQ devices  as current algorithms produce quantum circuits that increase in depth with increasing numbers of time-steps \cite{wiebe2011simulating}.  To limit the circuit size, we simulated a small AF Heisenberg model, eq.~\ref{AF_Heisenberg}, with only three spins on the IBM's Yorktown (\texttt{ibmqx2}) and Casablanca (\texttt{ibmq\_casablanca}) devices. 
\begin{figure}[ht!]
\centering
\begin{subfigure}{.5\textwidth}
  \centering
  \includegraphics[width=\linewidth]{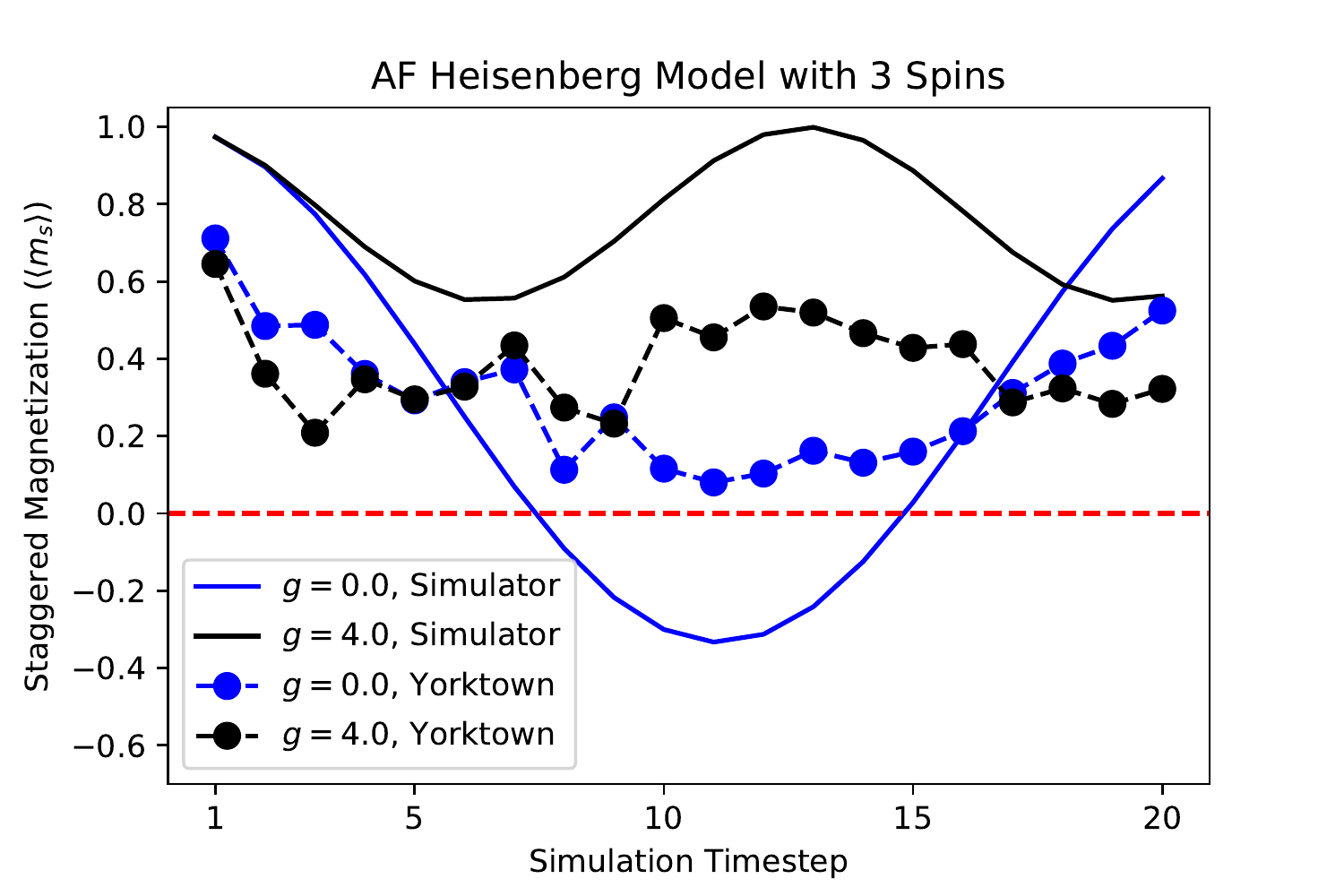}
  \label{fig:af_heisenberg_ibmq_yorktown}
  \caption{IBMQ Yorktown device}
\end{subfigure}%

\begin{subfigure}{.5\textwidth}
  \centering
  \includegraphics[width=\linewidth]{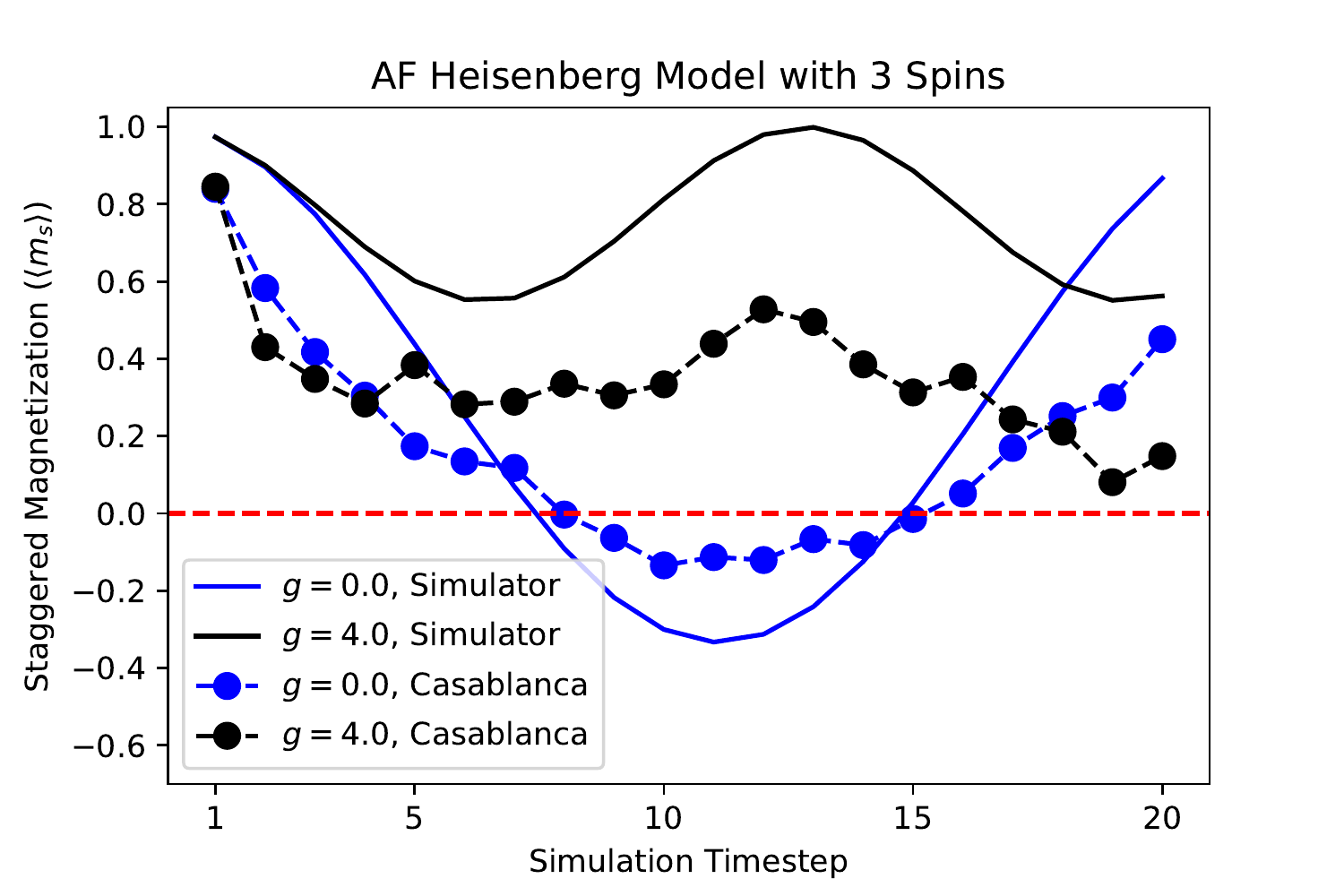}
  \label{fig:af_heisenberg_ibmq_casablanca}
  \caption{IBMQ Casablanca device}
\end{subfigure}
\caption{Results of simulation an AF model (eq.~\ref{AF_Heisenberg}) for a system with three spins using the code snippet in Fig.~\ref{fig:codesnippet} targeting the IBMQ's Yorktown (a) and Casablanca (b) devices. Each data point is an average of five runs of 8192 measurement shots each. The circuits are compiled and optimized using the QCOR compiler before submitting for execution. The Trotter time-step ($dt$) is 0.05.}
\label{fig:af_heisenberg_ibmq}
\end{figure}
\par
The simulation results from real quantum hardware for $g$ values of 0.0 and 4.0 are shown in Fig.~\ref{fig:af_heisenberg_ibmq}, where we can see the effects of gate-errors and qubit decoherence leading to a significant impairment of the measured staggered magnetization (circles) compared to the theoretical values (solid lines).  In particular, Fig.~\ref{fig:af_heisenberg_ibmq} demonstrates how the quality of the quantum hardware can affect simulation performance.  The Yorktown backend has considerably worse performance metrics than the Casablanca backend\footnote{Calibration data:\\
IBMQ Casablanca: Avg. CNOT Error:
1.165e-2, Avg. Readout Error:
2.069e-2, Avg. T1: 85.68 $\mu$s, Avg. T2:
78.5 $\mu$s.\\
IBMQ Yorktown: Avg. CNOT Error:
1.644e-2, Avg. Readout Error:
4.440e-2, Avg. T1: 50.95 $\mu$s, Avg. T2:
34.3 $\mu$s.}.  Specifically, compared to the Casablanca backend, Yorktown has a slightly higher two-qubit gate-error rate, nearly double the read-out error rate, and substantially lower qubit decoherence times.  While identical quantum circuits were run on the two machines, we see much better distinguishability between the results for the two values of $g$ in the results from Casablanca than those from Yorktown.

The staggered magnetization response to a quench for a simple three-qubit AF Heisenberg model in Fig.~\ref{fig:af_heisenberg_ibmq}, albeit noisy, illustrates non-trivial dynamics beyond that of decoherence (decaying to zero). Improvements in circuit construction (Trotter decomposition) and optimization, noise mitigation, and, most importantly, hardware performance (gate fidelity and qubit coherence) are required to scale up this time-domain simulation workflow for large quantum systems. 

\subsection{Variational Quantum Eigensolver}\label{VQE section}
\par
 As a second use case demonstration, we apply the Variational Quantum Eigensolver (VQE) algorithm to find the ground state energy of ${\rm H}_2$.  The VQE is a quantum-classic hybrid algorithm used to find a Hamiltonian's eigenvalues, where the quantum process side is represented by a parametrized quantum circuit whose parameters are updated by a classical optimization process\cite{peruzzo_variational_2014}.  The algorithm updates the quantum circuit parameters $\theta$ to minimize the Hamiltonian's expectation value $E_\theta$ until it converges. 

\begin{figure}[h!]
\centering
\begin{python}
from qcor import *
# Hamiltonian for H2
H = -0.22278593024287607*Z(3) + ... + \
    0.04532220205777769*X(0)*X(1)*Y(2)*Y(3) - \ 
    0.09886396978427353
# Defining the ansatz
@qjit
def ansatz(q : qreg, params : List[float]):
    X(q[0])
    ... 
    Rz(q[1],params[0])
    Rz(q[3],params[1])
    ... 
    Rz(q[3],params[2])
    ...
    H(q[3])   
    Rx(q[0], 1.57079)
# variational parameters
n_params = 3
# Create the problem model
problemModel = 
    QuaSiMo.ModelFactory.createModel(ansatz, 
                                  H, 
                                  n_params)
# Create the optimizer: spsa
optimizer = createOptimizer('nlopt',
                        {'algorithm':'spsa'})
# Create the VQE workflow
workflow = QuaSiMo.getWorkflow('vqe', 
                            {'optimizer': optimizer})
# Execute
result = workflow.execute(problemModel)
# Get the result
energy = result['energy']
\end{python}
\caption{Code snippet to learn the ground state energy of ${\rm H}_2$ by the VQE. For the sake of simplicity, we have omitted most of the terms in the Hamiltonian and ansatz.   }
\label{fig:codesnippet_vqe}
\end{figure}

\par
\begin{figure}[ht!]
    \centering
    \includegraphics[width = 0.5 \textwidth]{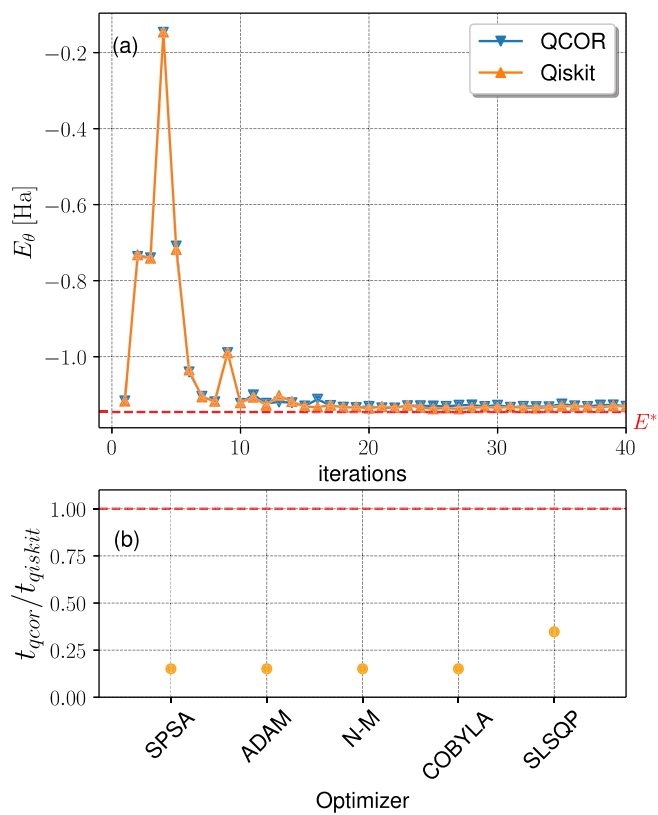}
    \caption{Energy ground state estimation for $H_2$ using VQE by using QCOR languange and Qiskit software. In panel (a), we present the VQE's learning path using the stochastic algorithm SPSA to find a quantum state with minimal energy. This plot shows how the energy $E_\theta$ approaches the exact value $E^* = -1.1456295 \ {\rm Ha}$ as the optimizer defines new quantum circuits, following the variational principle. In addition, panel (b), presents an execution time ratio between QCOR's VQE module and Qiskit's VQE module. The execution times ratio is below 1 (red dotted line) for different classical optimizers. The dots represent the 50th percentile of the execution time ratio from 51 independent runs of VQE. Optimizer labels correspond to Simultaneous Perturbation Stochastic Approximation (SPSA), ADAptive Moment estimation (ADAM), Nelder-Mead, also known as downhill simplex algorithm (N-M), Constraint Optimization By Linear Approximation (COBYLA), and the Sequential Least SQuares Programming (SLSQP).}
    \label{fig:H2vqe}
\end{figure}

The performance of the VQE algorithm, as any other quantum-classical variational algorithms \cite{benedetti_generative_2019}, depends in part on the selection of the classical optimizer and the circuit ansatz. The design scheme implemented in this work allows us to tune the VQE components to pursue better performance.  We present a listing of the code expressing this implementation in Fig. \ref{fig:codesnippet_vqe}, in which we define the different parameters of the VQE algorithm in a custom-tailored way. In the code snippet, \texttt{@qjit} is a directive to activate the QCOR just-in-time compiler, which compiles the kernel body into the intermediate representation, and \texttt{QuaSiMo.getWorkflow} is a utility function of the library to construct and initialize workflow objects of various types. In Fig. \ref{fig:H2vqe}(a), we present simulations considering the Simultaneous Perturbation Stochastic Approximation (SPSA) for ansatz updating. There, we show the energy as a function of the quantum circuits used in the learning process with a budget of 200 function evaluations with 5000 shots (experimental repetitions) per evaluation by using the QCOR's VQE module, \texttt{QuaSiMo.getWorkflow('vqe')}, and by using the Qiskit class \texttt{aqua.algorithms.VQE}~\cite{Qiskit-Aqua}. In both cases, we consider $(0.,0.,0.)$ as the initial set of parameters, therefore the energy values in the first iterations are similar, due to the SPSA hyperparameter's calibration.   
As it is expected, since we are using the same optimizer in both quantum simulations, there is not a relevant difference between the learning paths. However, the execution time in QCOR is shorter.  This feature is presented in Fig. \ref{fig:H2vqe}(b), where we show the execution time rate between QCOR and Qiskit as a function of the classical optimization algorithm.  To evaluate the typical execution time of each optimizer, we consider 51 runs of VQE towards the generation of $H_2$ quantum ground state.

\subsubsection{Symmetry reduction}
The presence of symmetries, such as rotations, reflections, number of particles, etc., in the Hamiltonian model allows us to map the model to a model with fewer qubits \cite{Bravyi2017}.  In $H_2$, we apply the QCOR's function \texttt{operatorTransform('qubit-tapering', H)} to reduce the four-qubit Hamiltonian, see Fig. \ref{fig:codesnippet_vqe}, to a one-qubit Hamiltonian model.  We present this implementation in Fig. \ref{fig:symredH2}, in which we transform the Hamiltonian introduced in Fig. \ref{fig:codesnippet_vqe} into a one-qubit model, and we redefine the anzats. In the output section of Fig. \ref{fig:symredH2}, we present the one-qubit model and the VQE's output that converge to a similar value of the four-qubit model. 
\begin{figure}[ht!]
\centering
  \begin{python}
# Now we taper Hamiltonian H2
H_tapered = operatorTransform('qubit-tapering', H)
# For the new Hamiltonian
# we must define an one qubit ansatz
@qjit
def ansatz(q : qreg, phi : float, theta : float):
    Rx(q[0], phi)
    Ry(q[0], theta)

 $ Output:
 $ Reduced Hamiltonian:
 $ (-0.328717,0) + (0.181289,0) X0 + (-0.787967,0) Z0
 $ with energy = -1.13727017466
 \end{python}
 \caption{Symmetry reduction of the Hamiltonian model. In this snippet, we transform the Hamiltonian model used in \ref{fig:H2vqe} to a one-qubit model following symmetry arguments. This feature is implemented in the function \texttt{operatorTransform('qubit-tapering', H)}. }
\label{fig:symredH2}
\end{figure}

\subsubsection{Fermion-qubit map}

\begin{figure}[ht!]
\centering
\begin{python}
from qcor import *
@qjit
def ansatz(q : qreg, x : List[float]):
    X(q[0])
    with decompose(q, kak) as u:
        from scipy.sparse.linalg import expm
        from openfermion.ops import QubitOperator
        from openfermion.transforms import \
            get_sparse_operator
        qop = QubitOperator('X0 Y1') \
            - QubitOperator('Y0 X1')
        qubit_sparse = get_sparse_operator(qop)
        u = expm(0.5j * x[0] * qubit_sparse).todense()
# Define the Hamiltonain
H = -2.1433 * X(0) * X(1) - 2.1433 * \
    Y(0) * Y(1) + .21829 * Z(0) - \
    6.125 * Z(1) + 5.907
num_params = 1
# Create the VQE problem model
problemModel = 
    QuaSiMo.ModelFactory.createModel(ansatz, 
                                     H, 
                                     num_params)
# Create the NLOpt derivative free optimizer
optimizer = createOptimizer('nlopt')
# Create the VQE workflow
workflow = QuaSiMo.getWorkflow('vqe', 
                    {'optimizer': optimizer})
# Execute the workflow 
# to determine the ground-state energy
result = workflow.execute(problemModel)
energy = result['energy']
\end{python}
\caption{Here we depict how to use OpenFermion operators to construct the state-preparation kernel (ansatz) for the VQE workflow.  We use SciPy \cite{2020SciPy-NMeth} and OpenFermion \cite{McClean2019OpenFermion} to construct the exponential of $(X_0Y_1 - Y_0X_1)$ operator as a matrix. The matrix will be decomposed into quantum gates by the QCOR compiler.}
\label{fig:codesnippet_vqe_fermionOp}
\end{figure}

An important feature included in the QCOR compiler is the fermion-to-qubit mapping that facilitates the quantum state searching in VQE. In Fig. \ref{fig:codesnippet_vqe_fermionOp}, we present an example of how to use OpenFermion operators \cite{McClean2019OpenFermion} in the VQE workflow. In that implementation, we define the ansatz by using Scipy and OpenFermion, the QCOR compiler decomposes the ansatz into quantum gates; we follow the same structure presented in Fig. \ref{fig:codesnippet_vqe} for the VQE workflow.

\subsection{Quantum Approximate Optimization Algorithm}
\label{QAOA section}
To further demonstrate the utility of QCOR we present an implementation of the quantum approximate optimization algorithm (QAOA) \cite{farhi2014quantum}.  QAOA translates a classical cost function into a quantum operator $H_C$ then uses a variational quantum-classical optimization loop to find quantum states that minimize the expectation value $\langle H_C\rangle$. The optimized quantum states are then prepared and measured to obtain bitstrings that correspond to classical solutions to the optimization problem. 

Figure \ref{fig:codesnippet_QAOA_MaxCut} shows example python code that uses QCOR simulations to find optimized quantum states with QAOA.  Opening lines define the number of qubits $n$ and construct the problem Hamiltonian $H_C$ for MaxCut on a star graph $S_n$.  The Hamiltonian is then used to create the \texttt{QuaSiMo} model and workflow to simulate $p$-step QAOA and return the expectation $\langle H_C\rangle$, similar to the previous VQE example of Fig.~\ref{fig:codesnippet_vqe}. 

\begin{figure}[t!]
    \centering
    \includegraphics[width=10cm]{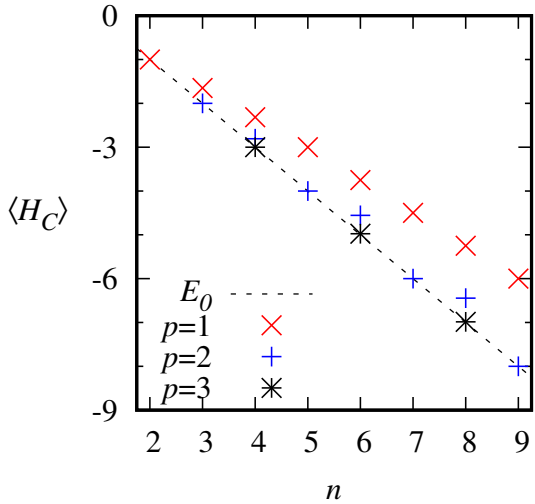}
    \caption{Optimized expectation values of the cost Hamiltonian $H_C$ for star-graph instances of unweighted MaxCut with QAOA.}
    \label{fig:QAOA star graphs}
\end{figure}

Figure \ref{fig:QAOA star graphs} shows optimized energies $\langle H_C\rangle$ computed in QCOR for star graphs with numbers of qubits $n=2,...,9$ at various QAOA depth parameters $p$.  Each time the program is run it begins with random initial parameters to be optimized, so we used several random starts for some of the graphs, keeping the smallest result as the $\langle H_C\rangle$ shown in the figure.  The QCOR results are in perfect agreement with the optimized standards \cite{lotshaw2021dataset} of Lotshaw {\it et al.~} \cite{lotshaw2021empirical}.  The results have interesting features, for example when $n$ is odd perfect ground state energies $\langle H_C\rangle = E_0$ are obtained at $p=2$, while when $n$ is even the results need $p=3$ to approach close to $E_0$.  The simplicity of the QCOR code and simulations make this an attractive route to studying interesting features like these in future research on QAOA.

\begin{figure}[ht!]
\centering
\begin{python}
from qcor import *
# QAOA for unweighted MaxCut
# Define a star graph Hamiltonian
n_qubits = 8
H = 0
for n in range(1,n_qubits):
    H += -0.5*(1-Z(0)*Z(n))
# create QuaSiMo model
problemModel = QuaSiMo.ModelFactory.createModel(H)
# Create the NLOpt derivative-free optimizer
optimizer = createOptimizer('nlopt')
# QAOA steps p
p = 3
# Quasimo workflow
workflow = QuaSiMo.getWorkflow('qaoa', 
                {'optimizer': optimizer, 
                'steps': p})
# Execute the workflow 
# to determine the energy expectation value
result = workflow.execute(problemModel)
energy = result['energy']
\end{python}
\caption{Code to find the expectation value of the cost Hamiltonian with QAOA, for a star-graph instance of the unweighted MaxCut problem.}
\label{fig:codesnippet_QAOA_MaxCut}
\end{figure}

\subsection{Quantum Imaginary Time Evolution Algorithm}
\label{QITE section}
To compute the ground-state energy of an arbitrary Hamiltonian, in addition to VQE as we have demonstrated in~\ref{VQE section}, there is another algorithm, so-called Quantum Imaginary Time Evolution (QITE)~\cite{motta2019determining, Yeter_Aydeniz_2021}, which does not require the use of an ansatz nor an optimizer. In QITE, we evolve the state through imaginary time $it \equiv \beta$ by applying the time-evolution operator $U = e^{-\beta H}$, which  minimizes the system energy exponentially. To do so, for each imaginary time-step $\Delta \beta$, we approximate the imaginary time evolution based on the results of the previous steps.

As shown in Fig.~\ref{fig:uml}, the QITE workflow is integrated into the \texttt{QuaSiMo} library. In this example, we demonstrate the use of QITE to find the ground state energy of a three-qubit transverse field Ising model (TFIM),
\begin{equation} 
    H = J_{z}\sum_{i=1}^{N-1} \sigma_{i}^{z}\sigma_{i+1}^{z} + h_x \sum_{i=1}^{N} \sigma_{i}^{x}
    \label{eq:tfim}
\end{equation}
where $J_z = h_x = -1.0$ and $N = 3$ is the number of spins (qubits).  

Fig.~\ref{fig:qite_code} is the code snippet to set up the \texttt{QuaSiMo} problem description and workflow for this problem. 
\begin{figure}[t!]
\centering
\begin{minted}[frame=single,framesep=5pt]{c++}
// Initial state preparation
__qpu__ void state_prep(qreg q) {
  // e.g., |100>
  X(q[0]);
}

// Create the Hamiltonian: 3-qubit TFIM
auto observable = -(Z(0) * Z(1) + Z(1) * Z(2) +
                    X(0) + X(1) + X(2));
// Construct the problem model
auto problemModel = 
       ModelFactory::createModel(state_prep, 
                                  &observable);

// Create the qsearch IRTransformation
auto qsearch_optimizer = 
    createTransformation("qsearch");
// QITE workflow: 20 steps with dbeta = 0.45
// Also, use qsearch to optimize 
// the propagating circuit.
auto workflow =
    getWorkflow("qite", 
                {{"steps", 20},
                {"step-size", 0.45},
                {"circuit-optimizer", 
                        qsearch_optimizer}});
// Execute
auto result = workflow->execute(problemModel);
\end{minted}
\caption{Code snippet (in C++) to find the ground-state energy of a three-qubit TFIM Hamiltonian using the QITE workflow. In this example, we run the QITE algorithm for a total imaginary time $\beta$ of 9.0 ($d\beta = 0.45$, 20 steps). Additionally, we use the QSearch algorithm from BQSKit to optimize QITE circuits during workflow execution. The \texttt{state\_prep} kernel is used to initialize the qubits into the desired initial state, e.g., $|100\rangle$ in this particular case. }
\label{fig:qite_code}
\end{figure}
The problem model is captured by a single Hamiltonian operator constructed by direct Pauli operator algebra. We also want to note that the Pauli operator algebra of \texttt{QuaSiMo} allows for hierarchical construction of the Hamiltonian, e.g., via for loops, suitable for generic Hamiltonians similar to the one described in eq.~(\ref{eq:tfim}).

For QITE workflow configurations, we set \texttt{step-size} to 0.45 and \texttt{steps} to 20, for a total imaginary time of $\beta = 9.0$ ($\hbar=1$). The system begins in different initial states by using a state-preparation circuit, as shown in Fig.~\ref{fig:qite_code}. Similar to the Python API, users can also use the utility function \texttt{getWorkflow} to retrieve an instance of the QITE workflow from the QCOR service registry with the name key "\texttt{qite}" as shown in Fig.~\ref{fig:qite_code}. 

The main drawback of QITE algorithm is that the propagating circuit size increases during the imaginary time-stepping procedure. To alleviate this constraint, especially for execution on NISQ hardware, \texttt{QuaSiMo}'s QITE workflow implementation support custom, externally-provided circuit optimizers that will be invoked during the algorithm execution to minimize the circuit depth. In this demonstration, we take advantage of the QSearch~\cite{qsearch} optimizer from BQSKit library, which is capable of synthesizing constant-depth circuits for a variety of common use cases including the TFIM model in~(\ref{eq:tfim}). For instance, in this example, the final QITE circuit (step = 20) has approximately 8,000 gates (3,000 CNOT gates), which is clearly beyond the capability of current NISQ devices. Thanks to QSearch, we can always re-synthesize a constant depth circuit with only 14 CNOT gates (87 gates in total) for any of the time steps, which is the theoretical lower bound~\cite{PhysRevA.69.062321}, $\frac{1}{4}(4^n - 3n - 1) = 13.5$, for CNOT gate count in three-qubit circuits. 
\begin{figure}
\begin{tikzpicture}
\begin{axis}[
      cycle list name=exotic,
      legend columns=3,
      xmin = 0, xmax = 20,
      xlabel = {QITE Steps}, 
      ylabel = {Energy ($\langle H \rangle$)}, 
      y label style={at={(axis description cs:-0.1, 0.5)},anchor=south},
      title = {QITE for three-qubit TFIM}]
\addplot
table[x=x, y=y] {
x  y
0  -2
1  -2.84427
2  -3.07795
3  -3.19728
4  -3.27945
5  -3.34246
6  -3.3895
7  -3.4233
8  -3.44682
9  -3.4628
10  -3.47345
11  -3.48047 
12  -3.48505
13  -3.48803
14  -3.48995
15  -3.49119
16  -3.49197
17  -3.49246
18  -3.49277
19  -3.49296
20  -3.49306
};
\addlegendentry{$|000\rangle$};

\addplot
table[x=x, y=y] {
x  y
0	0
1	-1.81066
2	-2.8419
3	-3.27279
4	-3.39792
5	-3.44234
6	-3.46267
7	-3.47405
8	-3.48105
9	-3.48551
10	-3.48838
11	-3.49022
12	-3.49139
13	-3.49214
14	-3.49262
15	-3.49291
16	-3.49308
17	-3.49319
18	-3.49324
19	-3.49326
20	-3.49326
};
\addlegendentry{$|100\rangle$};

\addplot
table[x=x, y=y] {
x  y
0	0
1	-1.57027
2	-2.70977
3	-3.27578
4	-3.42229
5	-3.46363
6	-3.47779
7	-3.48424
8	-3.48777
9	-3.48989
10	-3.49122
11	-3.49205
12	-3.49258
13	-3.49292
14	-3.49311
15	-3.49323
16	-3.49329
17	-3.49332
18	-3.49333
19	-3.49332
20	-3.4933
};
\addlegendentry{$|110\rangle$};

\addplot
table[x=x, y=y] {
x  y
0	-2
1	-3.2446
2	-3.47606
3	-3.4927
4	-3.49378
5	-3.49382
6	-3.49379
7	-3.49376
8	-3.49373
9	-3.4937
10	-3.49366
11	-3.49363
12	-3.49359
13	-3.49356
14	-3.49352
15	-3.49349
16	-3.49345
17	-3.49342
18	-3.49338
19	-3.49335
20	-3.49331
};
\addlegendentry{$(|000\rangle + |111\rangle)/\sqrt{2}$};
\addplot[mark=none, red, dashed, line width=1pt] coordinates {(0, -3.49396) (21, -3.49396)};
\end{axis}
\end{tikzpicture} 
\caption{QITE results for 3-qubit TFIM with different initial states, step size = 0.45. The red dashed line represents the true ground-state energy of -3.49396.}
\label{fig:qite_result}
\end{figure}
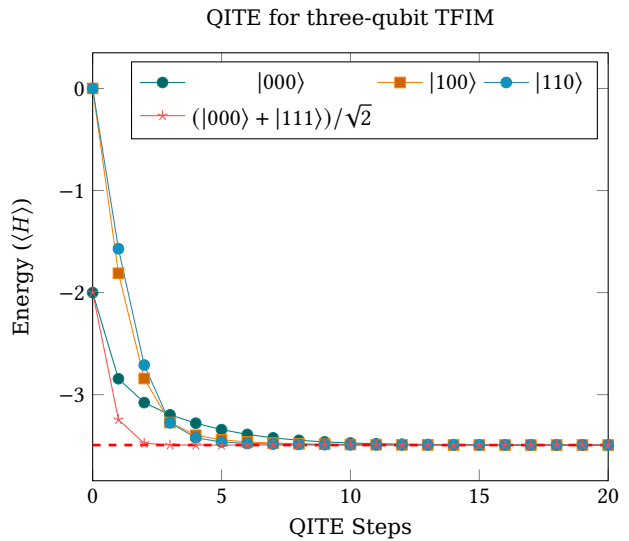
The results of the QITE workflow execution are shown in Fig.~\ref{fig:qite_result}, where we can see the energy value exponentially decays to the analytically computed ground state energy of -3.49396 for all initial states. 



\section{Conclusions}

We have presented and demonstrated a programming workflow for applications of quantum simulation that promotes common, reusable methods and data structures for scientific applications.
\par 
\tsh{We note that while the framework presented here is readily applicable to use cases in the NISQ era - with the \texttt{QuantumSimulationWorkflow} in particular being extendable to NISQ simulation algorithms such as VQE - the workflow is also extendable to universal algorithms. Code portability is ensured through intermediate representation that is then implemented for each backend according to specific APIs, which are accessible in the \texttt{QuaSiMo} library. This enables rapid porting of a simulation algorithm to multiple machines (write once, run everywhere paradigm). This is an ideal environment in which to construct both benchmarks and validation protocols, along with short depth quantum simulations that can quickly be run on multiple processors and directly compared.} 
\par
\tsh{While we maintain a focus on rapid prototyping of quantum simulation on today's NISQ devices, we note that the nature of the intermediate representation and the modular backend structure enable targeting fault tolerant devices as well. A fault tolerant (FT) architecture may be represented as an additional backend with encoding transpiler preprocessing. While this configuration is ideal for \texttt{QuaSiMo}'s unencoded qubit targets, we also note that the parent language QCOR, is capable of expressing fully quantum-error-correction-encoded algorithms as well. Therefore, we expect the framework to be extendable and to find use in workflows involving universal or FT applications in the future.}

\section*{Acknowledgments}
This work was supported by the U.S.~Department of Energy (DOE) Office of Science Advanced Scientific Computing Research program office Accelerated Research for Quantum Computing program. This research used resources of the Oak Ridge Leadership Computing Facility, which is a DOE Office of Science User Facility supported under Contract DE-AC05-00OR22725.

\bibliographystyle{iet}  
\bibliography{references}

\begin{thebibliography}{10}

\bibitem{aspuru2005simulated}
Aspuru.Guzik, A., Dutoi, A.D., Love, P.J., Head.Gordon, M.: `Simulated quantum
  computation of molecular energies', \emph{Science},  2005, \textbf{309},
  (5741), pp.~1704--1707

\bibitem{whitfield2011simulation}
Whitfield, J.D., Biamonte, J., Aspuru.Guzik, A.: `Simulation of electronic
  structure hamiltonians using quantum computers', \emph{Molecular Physics},
  2011, \textbf{109}, (5), pp.~735--750

\bibitem{cao2019quantum}
Cao, Y., Romero, J., Olson, J.P., Degroote, M., Johnson, P.D., Kieferov{\'a},
  M., et~al.: `Quantum chemistry in the age of quantum computing',
  \emph{Chemical reviews},  2019, \textbf{119}, (19), pp.~10856--10915

\bibitem{mcardle2020quantum}
McArdle, S., Endo, S., Aspuru.Guzik, A., Benjamin, S.C., Yuan, X.: `Quantum
  computational chemistry', \emph{Reviews of Modern Physics},  2020,
  \textbf{92}, (1), pp.~015003

\bibitem{yeter2020scattering}
Yeter.Aydeniz, K., Siopsis, G., Pooser, R.C.: `Scattering in the ising model
  using quantum lanczos algorithm', \emph{arXiv preprint arXiv:200808763},
  2020,

\bibitem{o2016scalable}
O'Malley, P.J., Babbush, R., Kivlichan, I.D., Romero, J., McClean, J.R.,
  Barends, R., et~al.: `Scalable quantum simulation of molecular energies',
  \emph{Physical Review X},  2016, \textbf{6}, (3), pp.~031007

\bibitem{lidar1999calculating}
Lidar, D.A., Wang, H.: `Calculating the thermal rate constant with exponential
  speedup on a quantum computer', \emph{Physical Review E},  1999, \textbf{59},
  (2), pp.~2429

\bibitem{bassman2020towards}
Bassman, L., Liu, K., Krishnamoorthy, A., Linker, T., Geng, Y., Shebib, D.,
  et~al.: `Towards simulation of the dynamics of materials on quantum
  computers', \emph{Physical Review B},  2020, \textbf{101}, (18), pp.~184305

\bibitem{kosugi2020linear}
Kosugi, T., Matsushita, Y.i.: `Linear-response functions of molecules on a
  quantum computer: Charge and spin responses and optical absorption',
  \emph{Physical Review Research},  2020, \textbf{2}, (3), pp.~033043

\bibitem{PhysRevLett.124.230501}
Lysne, N.K., Kuper, K.W., Poggi, P.M., Deutsch, I.H., Jessen, P.S.: `Small,
  highly accurate quantum processor for intermediate-depth quantum
  simulations', \emph{Phys Rev Lett},  2020, \textbf{124}, pp.~230501.
\newblock Available from:
  \url{https://link.aps.org/doi/10.1103/PhysRevLett.124.230501}

\bibitem{cirstoiu2020variational}
Cirstoiu, C., Holmes, Z., Iosue, J., Cincio, L., Coles, P.J., Sornborger, A.:
  `Variational fast forwarding for quantum simulation beyond the coherence
  time', \emph{npj Quantum Information},  2020, \textbf{6}, (1), pp.~1--10

\bibitem{google2020hartree}
Quantum, G.A., et~al.: `Hartree-fock on a superconducting qubit quantum
  computer', \emph{Science},  2020, \textbf{369}, (6507), pp.~1084--1089

\bibitem{mcclean2016theory}
McClean, J.R., Romero, J., Babbush, R., Aspuru.Guzik, A.: `The theory of
  variational hybrid quantum-classical algorithms', \emph{New Journal of
  Physics},  2016, \textbf{18}, (2), pp.~023023

\bibitem{romero2018strategies}
Romero, J., Babbush, R., McClean, J.R., Hempel, C., Love, P.J., Aspuru.Guzik,
  A.: `Strategies for quantum computing molecular energies using the unitary
  coupled cluster ansatz', \emph{Quantum Science and Technology},  2018,
  \textbf{4}, (1), pp.~014008

\bibitem{grimsley2019adaptive}
Grimsley, H.R., Economou, S.E., Barnes, E., Mayhall, N.J.: `An adaptive
  variational algorithm for exact molecular simulations on a quantum computer',
  \emph{Nature {C}ommunications},  2019, \textbf{10}, (1), pp.~1--9

\bibitem{tang2019qubit}
Tang, H.L., Barnes, E., Grimsley, H.R., Mayhall, N.J., Economou, S.E.:
  `qubit-adapt-vqe: An adaptive algorithm for constructing hardware-efficient
  ansatze on a quantum processor', \emph{arXiv preprint arXiv:191110205},
  2019,

\bibitem{farhi2014quantum}
Farhi, E., Goldstone, J., Gutmann, S.: `A quantum approximate optimization
  algorithm', \emph{arXiv:14114028},  2014,

\bibitem{motta2019determining}
Motta, M., Sun, C., Tan, A.T., O'Rourke, M.J., Ye, E., Minnich, A.J., et~al.:
  `Determining eigenstates and thermal states on a quantum computer using
  quantum imaginary time evolution', \emph{Nature Physics},  2019, pp.~ 1--6

\bibitem{Qiskit-Aqua}
Qiskit. `Qiskit aqua'. (Qiskit,  2021.
\newblock Available from: \url{https://github.com/Qiskit/qiskit-aqua}

\bibitem{Qiskit-Terra}
Qiskit. `Qiskit terra'. (Qiskit,  2021.
\newblock Available from: \url{https://github.com/Qiskit/qiskit-terra}

\bibitem{kottmann2020tequila}
Kottmann, J.S., Alperin.Lea, S., Tamayo.Mendoza, T., Cervera.Lierta, A.,
  Lavigne, C., Yen, T.C., et~al.: `Tequila: A platform for rapid development of
  quantum algorithms', \emph{arXiv preprint arXiv:201103057},  2020,

\bibitem{Orquestra}
Zapata. `Orquestra'. (Zapata,  2021.
\newblock Available from: \url{https://www.zapatacomputing.com/orquestra}

\bibitem{kandala2017hardware}
Kandala, A., Mezzacapo, A., Temme, K., Takita, M., Brink, M., Chow, J.M.,
  et~al.: `Hardware-efficient variational quantum eigensolver for small
  molecules and quantum magnets', \emph{Nature},  2017, \textbf{549}, (7671),
  pp.~242

\bibitem{hempel2018quantum}
Hempel, C., Maier, C., Romero, J., McClean, J., Monz, T., Shen, H., et~al.:
  `Quantum chemistry calculations on a trapped-ion quantum simulator',
  \emph{Physical Review X},  2018, \textbf{8}, (3), pp.~031022

\bibitem{mccaskey2019quantum}
McCaskey, A.J., Parks, Z.P., Jakowski, J., Moore, S.V., Morris, T.D., Humble,
  T.S., et~al.: `Quantum chemistry as a benchmark for near-term quantum
  computers', \emph{npj Quantum Information},  2019, \textbf{5}, (1), pp.~1--8

\bibitem{yeter2020practical}
Yeter.Aydeniz, K., Pooser, R.C., Siopsis, G.: `Practical quantum computation of
  chemical and nuclear energy levels using quantum imaginary time evolution and
  lanczos algorithms', \emph{npj Quantum Information},  2020, \textbf{6}, (1),
  pp.~1--8

\bibitem{qiskit-nature}
Qiskit. `Introducing qiskit nature'. (Qiskit,  2021.
\newblock Available from:
  \url{https://medium.com/qiskit/introducing-qiskit-nature-cb9e588bb004}

\bibitem{qcor2}
Nguyen, T., Santana, A., Kharazi, T., Claudino, D., Finkel, H., McCaskey, A.:
  `Extending c++ for heterogeneous quantum-classical computing', \emph{arXiv
  preprint arXiv:201003935},  2020,

\bibitem{osgi}
Marples, D., Kriens, P.: `{The Open Services Gateway Initiative: An
  introductory overview}', \emph{Communications Magazine, IEEE},  2002,
  \textbf{39}, (12), pp.~110--114

\bibitem{paper_repo}
`Source code repository'. (GitLab,  2020.
\newblock \url{https://code.ornl.gov/elwasif/qsa-workflow}

\bibitem{britt2017high}
Britt, K.A., Humble, T.S.: `High-performance computing with quantum processing
  units', \emph{ACM Journal on Emerging Technologies in Computing Systems
  (JETC)},  2017, \textbf{13}, (3), pp.~1--13

\bibitem{xacc1}
McCaskey, A.J., Dumitrescu, E.F., Liakh, D., Chen, M., Feng, W., Humble, T.S.:
  `A language and hardware independent approach to
  quantum{\textendash}classical computing', \emph{SoftwareX},  2018,
  \textbf{7}, pp.~245 -- 254.
\newblock Available from:
  \url{http://www.sciencedirect.com/science/article/pii/S2352711018300700}

\bibitem{marples2001open}
Marples, D., Kriens, P.: `The open services gateway initiative: An introductory
  overview', \emph{IEEE Communications magazine},  2001, \textbf{39}, (12),
  pp.~110--114

\bibitem{qcor1}
Mintz, T.M., McCaskey, A.J., Dumitrescu, E.F., Moore, S.V., Powers, S.,
  Lougovski, P.: `Qcor: A language extension specification for the
  heterogeneous quantum-classical model of computation', \emph{ACM Journal on
  Emerging Technologies in Computing Systems (JETC)},  2020, \textbf{16}, (2),
  pp.~1--17

\bibitem{trotter1959product}
Trotter, H.F.: `On the product of semi-groups of operators', \emph{Proceedings
  of the American Mathematical Society},  1959, \textbf{10}, (4), pp.~545--551

\bibitem{suzuki1976generalized}
Suzuki, M.: `Generalized trotter's formula and systematic approximants of
  exponential operators and inner derivations with applications to many-body
  problems', \emph{Communications in Mathematical Physics},  1976, \textbf{51},
  (2), pp.~183--190

\bibitem{barkoutsos2018quantum}
Barkoutsos, P.K., Gonthier, J.F., Sokolov, I., Moll, N., Salis, G., Fuhrer, A.,
  et~al.: `Quantum algorithms for electronic structure calculations:
  Particle-hole hamiltonian and optimized wave-function expansions',
  \emph{Physical Review A},  2018, \textbf{98}, (2), pp.~022322

\bibitem{o2020error}
O'Brien, T.E., Polla, S., Rubin, N.C., Huggins, W.J., McArdle, S., Boixo, S.,
  et~al.: `Error mitigation via verified phase estimation', \emph{arXiv
  preprint arXiv:201002538},  2020,

\bibitem{qcor_github}
`Qcor - c++ compiler for heterogeneous quantum-classical computing built on
  clang and xacc'. (GitHub,  2020.
\newblock \url{https://github.com/ORNL-QCI/qcor}

\bibitem{barmettler2010quantum}
Barmettler, P., Punk, M., Gritsev, V., Demler, E., Altman, E.: `Quantum
  quenches in the anisotropic spin-heisenberg chain: different approaches to
  many-body dynamics far from equilibrium', \emph{New Journal of Physics},
  2010, \textbf{12}, (5), pp.~055017

\bibitem{preskill2018quantum}
Preskill, J.: `Quantum computing in the nisq era and beyond', \emph{Quantum},
  2018, \textbf{2}, pp.~79

\bibitem{wiebe2011simulating}
Wiebe, N., Berry, D.W., H{\o}yer, P., Sanders, B.C.: `Simulating quantum
  dynamics on a quantum computer', \emph{Journal of Physics A: Mathematical and
  Theoretical},  2011, \textbf{44}, (44), pp.~445308

\bibitem{peruzzo_variational_2014}
Peruzzo, A., McClean, J., Shadbolt, P., Yung, M.H., Zhou, X.Q., Love, P.J.,
  et~al.: `A variational eigenvalue solver on a photonic quantum processor',
  \emph{Nature Communications},  2014, \textbf{5}, (1), pp.~4213.
\newblock Available from: \url{https://doi.org/10.1038/ncomms5213}

\bibitem{benedetti_generative_2019}
Benedetti, M., Garcia.Pintos, D., Perdomo, O., Leyton.Ortega, V., Nam, Y.,
  Perdomo.Ortiz, A.: `A generative modeling approach for benchmarking and
  training shallow quantum circuits', \emph{npj Quantum Information},  2019,
  \textbf{5}, (1), pp.~45.
\newblock Available from: \url{https://doi.org/10.1038/s41534-019-0157-8}

\bibitem{Bravyi2017}
Bravyi, S., Gambetta, J.M., Mezzacapo, A., Temme, K.: `Tappering off qubits to
  simulate fermionic hamiltonian', \emph{arXiv preprint arXiv:170108213},
  2017,

\bibitem{2020SciPy-NMeth}
Virtanen, P., Gommers, R., Oliphant, T.E., Haberland, M., Reddy, T.,
  Cournapeau, D., et~al.: `{{SciPy} 1.0: Fundamental Algorithms for Scientific
  Computing in Python}', \emph{Nature Methods},  2020, \textbf{17},
  pp.~261--272

\bibitem{McClean2019OpenFermion}
McClean, J.R., Sung, K.J., Kivlichan, I.D., Cao, Y., Dai, C., Fried, E.S.,
  et~al.: `Openfermion: The electronic structure package for quantum
  computers', ,  2019,

\bibitem{lotshaw2021dataset}
Lotshaw, P.C., Humble, T.S.. `{QAOA} dataset'. (,  2021.
\newblock Available from: \url{https://code.ornl.gov/qci/qaoa-dataset-version1}

\bibitem{lotshaw2021empirical}
Lotshaw, P.C., Humble, T.S., Herrman, R., Ostrowski, J., Siopsis, G.:
  `Empirical performance bounds for quantum approximate optimization',
  \emph{arXiv:2102:06813},  2021,

\bibitem{Yeter_Aydeniz_2021}
Yeter.Aydeniz, K., Siopsis, G., Pooser, R.C.: `Scattering in the ising model
  with the quantum lanczos algorithm', \emph{New Journal of Physics},  2021,
  \textbf{23}, (4), pp.~043033.
\newblock Available from: \url{https://doi.org/10.1088/1367-2630/abe63d}

\bibitem{qsearch}
Davis, M.G., Smith, E., Tudor, A., Sen, K., Siddiqi, I., Iancu, C.
\newblock `Towards optimal topology aware quantum circuit synthesis'.
\newblock In: 2020 IEEE International Conference on Quantum Computing and
  Engineering (QCE). (,  2020. pp.~ 223--234

\bibitem{PhysRevA.69.062321}
Shende, V.V., Markov, I.L., Bullock, S.S.: `Minimal universal two-qubit
  controlled-not-based circuits', \emph{Phys Rev A},  2004, \textbf{69},
  pp.~062321.
\newblock Available from:
  \url{https://link.aps.org/doi/10.1103/PhysRevA.69.062321}

\end{thebibliography}
\vspace{12pt}
\end{document}